# THE MAGNETITE AS ADSORBENT FOR SOME HAZARDOUS SPECIES FROM AQUEOUS SOLUTIONS: A REVIEW

Tanya M. Petrova [1], Ludmil Fachikov[2], Jordan Hristov[3]

**Abstract** – The review refers to the adsorption/desorption possibility of the magnetite, both natural and synthetic, with respect to hazardous species dissolved in aqueous solutions. The analysis stresses the attention on typical contaminants such as uranium, cadmium, cobalt, europium and arsenic. Most of the studies, performed so far, are on a laboratory scale without any attempts to be applied at larger either pilot or industrial scales. This especially invokes an analysis addressing what it would be when the scale of application increases beyond that of the laboratory flask. This point is of primary importance when fashionable nano-scale magnetite particles are used for sorption. The choice of magnetite is special because this mineral exhibits strong magnetic properties easily allowing creation of devices and processes for both upstream (adsorption) and downstream/ deposition processes such as fixed bed adsorption, magnetically stabilized beds, magnetic separation, and remote deposition of dangerous materials..

*Keywords*: magnetite, adsorption, hazardous species, aqueous solutions, separations

## I. Introduction

Environmental contamination by hazardous species is a wide spread problem, with sources of pollution arising from industrial activities and natural sources such as water contamination by arsenic, for example. Environmental contamination by heavy metals or radionuclides, for instance, is a wide spread problem. This review address adsorption processes performed by pure magnetite particles, both naturally occurring coarse ones (micron and millimeter size) and the modern magnetite nanoparticle. Composites based on resins, chitosan, and alginate, etc., commonly used also for separations are excluded from the analysis because in these composites, the magnetite component does not contribute to the adsorption but only ensures that the adsorbent beads behave as magnetic bodies. Magnetite, or iron ferrite ( $FeO \cdot Fe_2O_3$ ) is a naturally occurring mineral, but also can be easily prepared in the laboratory from solutions containing ferric and ferrous ions ([1], [2]). Moreover, the magnetite exhibits good adsorption characteristics with respect to a wide variety of species such as dissolved metals, particulate matter, and organic and biological materials, as an economical and environmentally inert material. The primary interest is in the adsorption properties of the magnetite surface as a prominent iron oxide and consequently the organization of the adsorption process, and the removal the adsorbents safely from the decontaminated solutions..

## II. The magnetite-basic properties

### II.1 The magnetite as a matter

Magnetite, one of the important iron ores, can be found everywhere in nature, in igneous and metamorphic rocks. The presence of magnetite in nature is often a result of biological processes, but it can also have a lithogenic origin [1] It is found also in ocean floor, soils, rocks, meteorites, atmospheric dust, bacteria and other living organisms. It is also a common corrosion product of iron and steel. Magnetite ( $Fe_3O_4$ ) is a commonly found in the environment and can form via several pathways, including biotic and abiotic reduction of ferric iron $Fe^{3+}$ oxides and the oxidation of ferrous iron $Fe^{2+}$ and iron metal ( $Fe^O$ ).Most of the Fe oxides, such as goethite, hematite, lepidocrocite, and maghemite are semiconductors, whereas magnetite exhibits properties closer to that of a metal. Magnetite has also been shown to reduce several contaminants in laboratory studies, such as carbon tetrachloride ( $CCl_4$ ) [3], hexavalent chromium $Cr^{6+}$ [4]; [5], hexavalent uranium ( $U^{6+}$ ), and several other compounds .

The magnetite crystal structure is inverse spinel with a unit cell consisting of 32 oxygen atoms in a face-centered cubic structure and a unit cell edge length of $0.839\,nm$ [6], responsible for both the magnetic property and the colour of the mineral. The black colour is due to the intervalence charge transfer between $Fe^{2+}$ and $Fe^{3+}$ in its crystal structure. The magnetic property in magnetite also stems from the $Fe^{2+}$ and $Fe^{3+}$ atoms being in tetra and octahedral sites as described above. The spins on the tetrahedral and octahedral sites are antiparallel and the magnitudes of the two types of spins are unequal [6].In oxidizing atmosphere, magnetite is oxidized to maghemite or hematite, namely:

$$4Fe_3O_4 + O_2 \rightarrow 6Fe_2O_3 \qquad (1)$$

In a reducing atmosphere, for instance carbon, it can be reduced to wustite or iron:

$$Fe_3O_4 + C \rightarrow 3FeO + CO \qquad (2a)$$
$$Fe_3O_4 + 4C \rightarrow 3Fe + 4CO \qquad (2b)$$

Magnetite is usually characterized by several methods providing information about chemical properties and crystal structure, among them: infrared spectroscopy, ultraviolet and visible spectroscopy ($UV-VIS$), X-ray diffractometry ($XRD$), thermal analysis, magnetometry, and Mössbauer spectroscopy. The surface and morphology can be investigated by X-ray photoelectron spectroscopy ($XPS$) and TEM or SEM microscopic techniques, as well.

Additionally to the natural resources, a number of methods have been developed to synthesize magnetite with specific characteristics, among them : co-precipitation [7], oxidation of ferrous hydroxide $Fe(OH)_2$ [8], reduction of hematite at $400^oC$ and in $H_2/Air$ as reducing atmosphere, alkaline hydrolysis of ferrous sulphate followed by calcination at $500^oC$, and oxidation of ferrous sulphate at alkaline condition by potassium nitrate [9]. Magnetite has a relatively low solubility in pure water where the pH is close to point of zero charge (**PZC**) for magnetite (established as $pzc = 6-6.8$) [6], however, solubility could be increased by changing $pH$ and/or in presence of complexing agents. Microorganisms can also dissolve magnetite [10]. In aqueous systems, iron oxides act as Lewis acids and adsorb water or hydroxyl groups (singly, doubly and triply coordinated $Fe$ atoms). Moreover, two $OH$ groups can coordinate to one iron atom and this depends on the crystal structure and morphology of the iron oxide.

*II.2 Adsorption issues*

One important issue in surface adsorption studies of minerals is that the adsorption process occurs only at the surface and hence, it is advantageous from an experimental point of view if the mineral has a high surface area. Focusing on adsorption studies of hazardous materials onto magnetite only, this review tries to find the appropriate models to describe the adsorption behaviour at the surface and many factors affecting it. Moreover, this analysis will allow us encompass the current situation and to draw some common tendencies in application of magnetite as successful adsorbent for wastewater decontamination and mitigation of the effect of polluted hazardous substances.

*II.2.1. Adsorption Isotherms*

Probably, the most often used model to describe the adsorption onto magnetite is the Langmuir isotherm originally derived for the adsorption of gases onto solid surfaces. It is based on a model with the basic assumptions: i) There is only a monomolecular layer of adsorbed molecules: ii). The equilibrium is characterized by the fact that the rates of adsorption ($r_{sorption} = k_s(q_{max} - q_i)c_i$) and desorption ($r_{desorption} = k_{des}q_i$) are equal. With $b = k_{des}/k_s$ this assumption leads to

$$q_i = q_{max}c_i\left(\frac{1}{k_{des}/k_s + c_i}\right) \Rightarrow \Rightarrow q_i = q_{max}\frac{c_i}{b+c_i} \qquad (3)$$

By means of the reverse ratio $K = k_s/k_{des}$ the Langmuir isotherm can be expressed in one frequently used form

$$q_i = q_{max}\frac{Kc}{1+Kc} \Rightarrow \Theta = \frac{Kc}{1+Kc} \qquad (4)$$

The model links the fraction of the surface sites covered by the adsorbate $\Theta = A/A_{max}$, the bulk concentration of the adsorbate $c$ and the affinity constant $K$. $A$ is the amount of the adsorbate that is adsorbed at equilibrium concentration $c$; $A_{max}$ is the amount that is necessary to cover the surface completely by a monolayer of adsorbate . -

The Langmuir relationship has two limiting cases:
- For very low concentrations the denominator tends to the unit ($Kc \ll 1$) and the isotherm becomes linear $q_i = q_{max}Kc$
- At high concentrations we get $q_i = q_{max}$, i.e. the sorption capacity has reached a maximum. The entire surface is covered with a mono-molecular layer, and a further increase in the solution concentration does not increase the amount adsorbed.

Commonly, the adsorption data are fitted by the Freundlich equation, which entire empirical relationship, namely

$$q = K_dC^{\frac{1}{n}} \qquad (5)$$

The Freundlich isotherm does not exhibit a maximum number of adsorption sites. It fits well data without physical meaning because eq. (5) teaches that the adsorbed amount grows without restriction as the bulk concentration $C$ increases. However, despite this formal drawback, this power-law relationship allows the surface heterogeneities to be taken into account, in away more general than the Langmuir isotherm. Thus, the equation applies very well to solids with heterogeneous surface properties and generally for heterogeneous solid surfaces [11]. The Freundlich equation was used for fitting data related to adsorption of cations on iron oxides [6].

The Temkin equation, is one that uses a logarithmic relationship between the amount adsorbed and concentration in solution onto non-uniform surfaces, predominately to chemisorption, namely

$$q = a \ln C \qquad (6)$$

where $a$ is a constant. This equation accounts the heterogeneities in the adsorbent surfaces.

*II.2.2. Sorption Kinetics*

The sorption process is not an instantaneous process because the species have to diffuse from the solution to the surface of the sorbent, and then to the internal surface areas and the overall rate of this approach to equilibrium defines the sorption kinetics. The rate of sorption is usually limited by mass transfer in the liquid phase (affected by the hydrodynamics) and depends on the properties of the sorbate and sorbent. Solid-liquid sorption processes are dominated either by the transport of solute molecules from the bulk solution across a film surrounding the adsorbent particles (film diffusion) or by the transport of the sorbate from the particle surface into interior sites by diffusion in the porous system of the adsorbent. Generally, both mechanisms affect the adsorption. Depending on the conditions of sorption one transport step may be much slow than the other one and, hence exclusively control the adsorption. Film diffusion is considered as the rate-controlling step in systems with poor mixing, small concentration of solute, small particle size and high affinity of the sorbate for the adsorbent. Internal diffusion dominates in systems with high concentration of sorbate, fast superficial overflow, and large particle size of adsorbent and low affinity of the sorbate for the adsorbent. In the context of the latter comments, when the solute concentration is small and the adsorbent size matches the nano-scale range, the film diffusion is the process expected to be the controlling step.

*II.2.2.1. Alternative models for describing sorption kinetics*

There exist sorption systems in which the film model does not work well in description of the experimental data. To overcome this problem, some alternative approaches describing the sorption kinetics of metal ions onto have to be applied, among them:

- **Pseudo-first-order chemical reaction** [12]. The pseudo-first order equation was applied for describing the adsorption of liquid-solid systems based on solid capacity [13]. Similar to that of a first-order chemical reaction, with the change of loading being proportional on the difference between the equilibrium loading, $q_e$, the relationship was suggested as

$$\frac{dq_t}{dt} = k_1(q_e - q_t), \ q_t(t=0) = 0 \qquad (7a)$$

$$\log(q_e - q_t) = \log q_e - \frac{k_1 t}{2.303} \qquad (7b)$$

The parameter $k_1(q_e - q_t)$ does not represent the number of salable sites and $\log q_e$ is adjustable parameter not equal to the intercept $\log(q_e - q_t)$ at $t = 0$, and in to fit correctly the experimental data the equilibrium sorption capacity, $q_e$ has to be known.

- **Pseudo-second-order approach** [12] assumes that rate of sorption is proportional to the square of the number of unoccupied sites of the adsorbent.

$$\frac{dq_t}{dt} = k_2(q_e - q_t)^2, \ q_t(t=0) = 0 \qquad (8a)$$

$$\frac{1}{q_e - q_t} = \frac{1}{q_e} + k_2 t \Rightarrow q_t = \frac{t}{1/k_2 q_e^2 + t/q_e} \Rightarrow$$
$$\Rightarrow \frac{t}{q_t} = \frac{1}{h} + \frac{t}{q_e} \qquad (8b)$$

Here, $h = k_2 q_e^2$ representing the initial sorption rate. the plot of vs. $t/q_t$ vs $t$ is an almost linear relationship that gives $q_e$, $h$ and $k_2$ without need of any parameters to be known in advance.

- **The Elovich equation [12]**, initially developed for description of kinetic of heterogeneous surfaces chemisorption of gases on solids, was successfully applied to liquid-solid systems, mainly for adsorption f inorganic substances on solid surfaces, namely

$$q_t = \left(\frac{1}{\beta}\right) \ln \alpha \beta + \frac{1}{\beta} \ln t \qquad (9)$$

The parameters $\alpha$ and $\beta$ are constants which ca be determined from the linear plot $q_t$ vs. $\ln t$, with a slope of $1/\beta$ and an intercept of $(1/\beta) \ln \alpha \beta$.

*II. 3. Point of Zero Charge (PZC)*

Surface adsorption operates through $Fe-OH$ groups at the surface of $Fe$ oxides and results from the completion of the ligand shell of surface $Fe$ atoms. These groups attain negative or positive charge by dissociation $(\equiv FeOH \rightarrow \equiv FeO^- + H^+)$ or association $(\equiv FeOH + H^+ \rightarrow FeOH_2^+)$ of protons. *The surface charging is a pH-dependent phenomenon*. There is a substantial difference between the dry and wet surface chemistry, especially for metal oxides dispersed in water [14]. The point is how ageing is expressed directly through changes in the aqueous interfacial properties of magnetite nanoparticles . In an aqueous medium, $pH$ and ionic strength dependent charges develop on amphoteric surface hydroxyls $(\underline{Fe} \cdot OH)$. The following protonation and deprotonation reactions can take place on the surface of iron oxide particles:

$$\underline{Fe} \cdot OH + H^+ \leftrightarrow \underline{Fe} \cdot OH_2^+ \quad (10)$$

$$\underline{Fe} \cdot OH \leftrightarrow \underline{Fe} \cdot O^- + H^+ \quad (11)$$

or

$$\underline{Fe} \cdot OH + OH^- \leftrightarrow \underline{Fe} \cdot O^- + H_2O \quad (12)$$

The $pH$ at which the surface concentrations of $FeOH_2^+$ and $FeO^-$ groups are equal, is the so-called point of zero charge ($PZC$). It is generally around $pH\,8-9$ for all iron oxides [15]. The magnitude of the $pH$ dependent charge is proportional to $(pH_{PZC} - pH)$ and to the ionic strength of the equilibrium solution. The charge will be in the range of several $\mu mol$ per $m^2$ of the oxide surface. According to the IUPAC recommendation [16] $H$ and $OH$ ions are considered to be the potential-determining ions. When surface charge development occurs by direct proton transfer from the aqueous phase, the surface charge density ($\sigma_{O,H}$) and surface potential ($\Psi_0$) can be defined analogously to the Nernstian surfaces:

$$\sigma_{O,H} = F\left(\Gamma_{H^+} - \Gamma_{OH^-}\right) \quad (13)$$

$$\Psi_0 = \left(\frac{RT}{F}\right)\ln\frac{[H^+]}{[H^+]_{PZC}} = 2.3\left(\frac{RT}{F}\right)(pH_{PZC} - pH) \quad (14)$$

where $\Gamma_i$ is the surface excess concentration of species $i$, $R$ is the gas constant, $T$ is temperature and $F$ is the Faraday constant.

Characterization of the change in the pH-dependent surface charging of magnetite during longer storage is of great practical importance in terms of application of water-based magnetic fluids [17]. The $pH$ of $pzc$ is characteristic of each metal oxide in an aqueous medium. Several experimental data of **pzc** for iron oxides are available in the literature, the values being between 3.8 and 9.9 for magnetite [17],[18], that matches to some extent the above mentioned range $pH\,8-9$. According to Cornell [6] the $PZC$ of magnetite is around 6 (see below, too) and the surface charge is close to neutral. Sun et al. [2] studied the surface characteristic of magnetite in aqueous suspension using potentiometric titration and concluded that the surface of magnetite contains surface hydroxyl groups which are protonated at $pH$ below the $PZC$ of magnetite (around 6), and deprotonated above this $pH$:

$$\equiv Fe(II,III)OH + H^+ \rightarrow Fe(II,III)OH_2^+,\; pH < PZC \quad (15)$$

$$\equiv Fe(II,III)OH - H^+ \rightarrow Fe(II,III)O^-,\; pH > PZC \quad (16)$$

Usually, the $pzc$ of magnetite is assumed about 6.4 at room temperature [19], but it also depends strongly on the temperature, the way the magnetite is synthesized, and the conditions under which the measurements are taken. In general, the $pzc$ will vary with the particle concentration, the ionic strength of the medium [20]. The electrostatic field developing around magnetite particles under acidic and alkaline conditions far from the $pH$ of $pzc$ can prevent particle aggregation due to the repulsion of overlapping electric double layers of approaching particles with similar charges [21]. The $pzc$ was also determined to be at $pH\,7.9 \pm 0.1$ for magnetite [6], thus matching the above-mentioned range [6].

### III. Adsorbent size

There exists an astonishing plethora of articles published on magnetite and application for wastewater treatment. The following analysis cannot encompass all published reports so we focus the attention on two groups based on the size of the magnetite particles used: **coarse particles** and **nano-particles**. This approach is based on some differences in their properties and modes of application to adsorption processes, among them:

- The **coarse magnetite particles** dominated in the literature before the era of the modern nano-technology. The coarse magnetite-based adsorbents exhibit surface areas comparable to all other natural metal oxide used for the same purposes, i.e. water decontamination. The natural rock magnetite and the magnetite sands with some surface contaminations are the principle sources of coarse particles.

- The **magnetite nano-particles**, in the last 10 years and at present are the only ones used for sorption processes in various aspects. The interest in this direction is mainly driven by either the common fashion in the nano-science or the high surface area exhibited by such particles even tough new unsolved problems concerning the process performance at scale beyond the laboratory flask as well separation/ deposition operations emerge.

### III.1. Coarse particles

Coarse magnetite particles are sometimes found in large quantities in beach sand. Such black sands (mineral sands or iron sands) are found in various places, such as California and the west coast of New Zealand. The magnetite sand is carried to the beach via rivers from erosion and is concentrated via wave action and currents. Such sands are of micron-size, flow well (as dry granular media) in tubes, vessels, and undergo fluidization without problem by either gas or liquids. Silicates, carbonates and other traces that affect both the color and the surface sorption properties commonly contaminate the grain surfaces. Desired fractions can be easily obtained by sieving.

### III.2. Nano-Particles

In the last two decades, nanomaterials have received considerable attention due to their small particle size, large surface area, low cost and ease of preparation. Some of the major attractive features are related to large and controllable surface area, low-cost production, non-toxicity to some extent, and ability to work in classical process with mechanical mixing of the contaminated aqueous media. Especially, to the wastewater treatment, the nanoparticles offer great applicability but at the same time challenge development of new devices and operating conditions different from those with classical coarse adsorbents. These particles are easily separable from $(\sim 98\%)$ water solution by the high-gradient magnetic separation (HGMS). Especially to magnetite-based nanoparticles, extremely small size of about $10\,nm$ can be easily obtained thus providing high contact surface area. The high surface area to some extent compensates the increase in the mass transfer resistance due to the stagnant liquid layer attached to the solid surface.

### III.2.1. Advantages of magnetic nano-particles

Magnetic nanoparticles display the phenomenon of super-paramagnetism, not keeping magnetized after the action of magnetic field, offering advantage of reducing risk of particle aggregation. First, they have sizes that place them at dimensions comparable to those of a virus ($20 \pm 500\,nm$ or proteins ($5 \pm 50\,nm$ [22]. The magnetic nanoparticles used in bio-applications are usually made from biocompatible materials such as magnetite ($Fe_3O_4$ Fe3O4) for which susceptibility is large [23].

### III.2.1.1. Synthetic magnetite nanoparticles

- **Co-precipitation from solution.** This method may be the most promising one because of its simplicity and productivity. The common way of magnetite synthesis is the alkaline hydrolysis of $Fe(II)$- and $Fe(III)$-salts. The size of formed particles depends on the relative oversaturation of solution; formation of nanoparticles is expected at very low and very high concentration according to the Weimarn rule. Single magnetic domains with size below $\sim 10\,nm$ can be developed under appropriate hydrolysis conditions. Tailored magnetite nanoparticles at a scale of $1.5-13\,nm$ can be developed by special control of both $pH$ and ionic strength in the co-precipitation medium. However, the smaller the particles, the less stable systems form in colloidal point of view because they tend to aggregate. The difficulty in preparing $Fe_3O_4$ magnetic nanoparticles by chemical co-precipitation is the tendency of agglomeration of particles because of extremely small particle size leading to great specific surface area and high surface energy, consideration of effect of alkali, emulsifier, and reaction temperature are the decisions making factors of final product [24], [25].
- **High-temperature decomposition of organic precursors** in the presence of hot organic surfactants has yielded markedly improved samples with good size control, narrow size distribution and good cristalinity of individual and dispersible magnetic iron oxide nanoparticles [26].
- **Microemulsions.** Water-in-oil ($W/O$) microemulsions systems, a fine micro droplets of the aqueous phase trapped within assemblies of surfactant molecules dispersed in a continuous oil phase [22]. The surfactant-stabilized micro cavities (typically in the range of 10 nm) provide a confinement effect that limits particle nucleation, growth, and agglomeration [27].
- **Polyols.** Fine metallic particles can be obtained reduction of dissolved metallic salts and direct metal precipitation from a solution containing a polyol [28].
- **Aerosol/vapor methods.** Spray and laser pyrolysis are excellent techniques for the direct and continuous production of well-defined magnetic nanoparticles under exhaustive control of the experimental conditions [22].

### III.2.2. Magnetic properties

The magnetic response of a specific mineral is largely dependent on its magnetic susceptibility. ($\chi$), defined as a ratio of the induced magnetization ($M$) to the applied magnetic field ($H$). In ferri- and ferromagnetic materials, magnetic moments orient themselves parallel to $H$, resulting in ordered magnetic states. The magnetic susceptibilities of these materials depend on their temperature, external field $H$ and atomic structures. At small sizes (in the order of tens of nanometers), ferri-or ferro-magnetic materials, become a single magnetic domain and therefore maintain one large magnetic moment. At sufficiently high temperatures (i.e., blocking temperature), the thermal energy is sufficient to induce free rotation of the particle resulting in a loss of their net magnetization in the absence of an external field [20]. Lack of remnant magnetization when the external field is turned-off allows the particles to maintain their colloidal stability and avoids aggregation. The coupling interactions in single magnetic domains result in much higher magnetic susceptibilities than paramagnetic materials. The coercive force of ferromagnetic particles decreases rapidly when the particle size decreases below the superparamagnetic size [24]. At room temperature, a coercive force of a few $Oe$ was observed for ultrafine magnetite particles. The saturation magnetization ($Ms$) per net weight of fine $Fe_3O_4$ particles (about $54\,emu/g$) is also smaller than that of bulk magnetite ($92\,emu/g$) [24].

## IV. Adsorption-and Desorption: *An Overview*

The overview addresses the sorption of radioactive contaminates and arsenic as primary hazardous substances affecting the human health. Due the limited format of the article, only the principle and most relevant studies are encompassed. The readers looking for more details may refer to the literature sources quoted in these articles.

### IV.1. Radioactive Contaminants

Our interest in the study of adsorption from solution lies in its possible role in the transport of radioactivity in waters. In particular, the transport of cobalt adsorbed on corrosion product oxides seems to be an important factor in the growth of radiation fields of both nuclear reactors and deposits of used fuel. In the absence of external magnetic fields, magnetite readily adsorbs numerous metal species including actinide elements and metals [31]. In the presence of an external magnetic field, a synergistic effect was observed in using supported magnetite in a fixed-bed for removal of plutonium and americium [31]. According to Cotten et al. [32] the low adsorption properties of magnetite for metal ions can be overcome if the solution can be altered to place the metals in a solid form such as a colloid. Other literature citations [33] have reported varying degrees of contaminant removal with magnetite both in and out of an external magnetic field. Petkovic and Milonjic [33] performed the adsorption of cesium on magnetite in batch experiments ($pH$ between $7.6$ and $10.4$) well described by the Langmuir adsorption isotherms.

### IV.1.1. Cobalt

Milonjic and Ruvarac [34] reported the adsorption of cesium ($Cs^+$), cobalt ($Co^{2+}$ as less than $5\%\,wt$) and cerium ($Ce^+$) by magnetite at pH of 2.2 by batch experiments. In this context, the adsorptive properties of magnetite with respect to alkali metal ions were also investigated in batch experiments [35]. Tamura et al. [36] reported adsorption of the long-lived isotope $^{60}C0^{2+}$ onto spherical magnetite particles, a corrosion product from the inner surfaces of the cooling pipe systems in the nuclear power stations. The adsorption of heavy metals by oxides may be expressed as [36]:

1) The exchange of metal ions $M^{m+}$ with protons at the surface hydroxyl groups - $MOH$ (i.e. surface complexation)

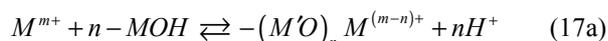

$$M^{m+} + n-MOH \rightleftarrows -(M'O)_n M^{(m-n)+} + nH^+ \qquad (17a)$$

2) The hydrolysis is followed by adsorption

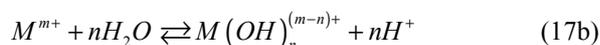

$$M^{m+} + nH_2O \rightleftarrows M(OH)_n^{(m-n)+} + nH^+ \qquad (17b)$$

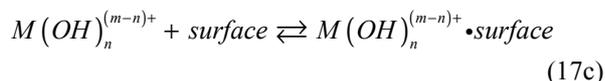

$$M(OH)_n^{(m-n)+} + surface \rightleftarrows M(OH)_n^{(m-n)+} \bullet surface$$
$$(17c)$$

3) Electrostatic bonds between the metal ions and the oxide surfaces

The hydrolysis is considered to weaken the interaction between the metal ion and water of hydration that facilitates the approach of the metal ions to the surface. The batch experiments of Tamura et al. [36] reveal that the amount of $Co^{2+}$ adsorbed on the magnetite depends on the $pH$ of the medium (See Fig.1). The adsorption equilibria was explained only in terms of chemical interaction of $-FOH$ groups with the surface.

Cotten and Navratil [32] observed the adsorption of cobalt in a static field of an neodymium-iron-boron permanent magnet irradiating a fixed bed of magnetite ($10\,mm\,I.D.$ and $50\,mm$ depth) mixed with glass spheres ($50/50$ by weight). The plots in Fig. 2a show clearly the effect of the magnetite on the sorption performance of the magnetite, then enhanced by the external field magnetization (Fig. 2b). According to the authors, the ionic sorption is small in comparison to the particle sorption capacity: recall the cobalt is also magnetic, so the external field boosts the magnetic attraction to the magnetite surfaces.

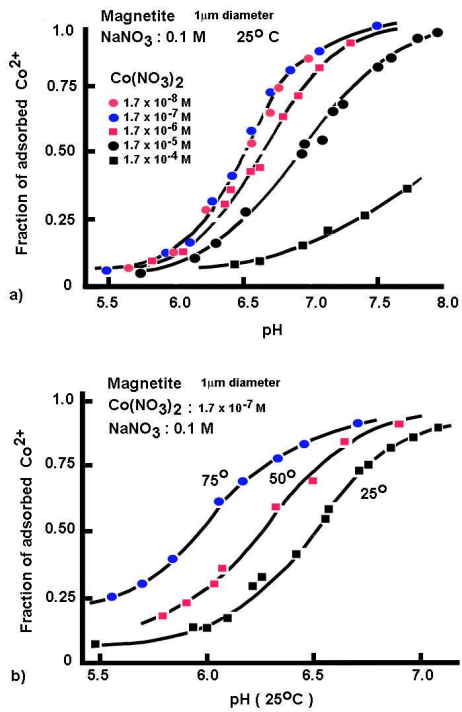

Fig.1. Adsorption of $Co^{2+}$ onto magnetite Adapted from Tamura et al. [36]. a) Fraction adsorbed as function of $pH$ at various initial concentrations of $Co^{2+}$ b). Fraction adsorbed as function of $pH(25^oC)$ at a constant concentration of $Co(NO_3)_2 = 1.7 \times 10^{-7} \, mol/dm^3$ and ionic strength $0.1 \, mol/dm^3$ at 3 different temperatures

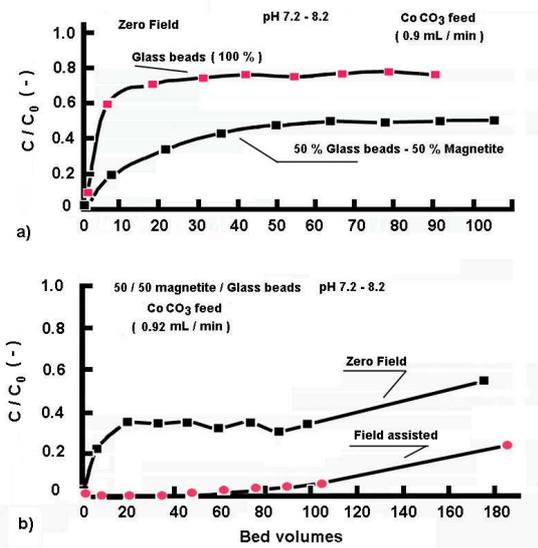

Fig. 2. Adsorption of $Co^{2+}$ onto magnetite/glass admixtures. Adapted from Cotten ad Navratil [32].
a) Comparative experiments on the effect of the magnetite content on the adsorption efficiency without magnetic field.
b) Magnetic field assisted adsorption in 50/50 Magnetite/Glass admixture bed

Tewari et al. [37] reported absorption of $Co^{2+}$ by oxides ($Fe_3O_4$, $Al_2O_3$ and $MnO_2$) as a function of the solute concentration, $pH$ and temperature. In all cases the adsorption increases with increase in $pH$ with the range from $5.0$ to $7.5$. Beyond this rage the adsorption becomes masked by precipitation of $Co(OH)_2$ that leads to loss of $Co^{2+}$ from the solution. The Langmuir isotherm was used to fit the data of the adsorption in the form.

$$\theta = \frac{ac_{eq}}{1 + ac_{eq}} \quad (18)$$

In (18) $a$ is a constant related to the intensity of adsorption ($a = K$ see eq. 4). In this context, the heat of adsorption was calculated by

$$\frac{c_{eq}}{w} = \frac{c_{eq}}{M_{sat}} + \frac{1}{aM_{sat}} \quad (19)$$

The saturation amount adsorbed $M_{sat}$, i.e. the capacity of adsorption for magnetite was estimated as $3.7 \times 10^{-2} \, \mu g/cm^2$ at $pH = 6.5$, while under the same condition $M_{sat}(Al_2O_3) = 6.6 \times 10^{-4} \, \mu g/cm^2$ and $M_{sat}(MnO_2) = 4.2 \times 10^{-2} \, \mu g/cm^2$ at $pH = 6.0$. Hence, the adsorption capacities of the magnetite and the $MnO_2$ are of the same order of magnitude. Additionally, the heat of adsorption with magnetite as adsorbent was established as $6.6 \, kcal/mol$. The heat of adsorption for magnetite at different values of the amount adsorbed is shown in Fig. 3.

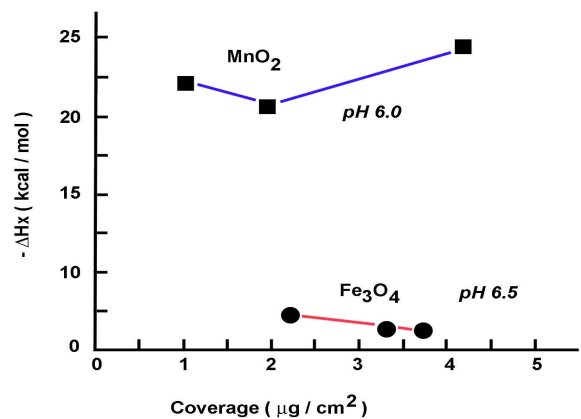

Fig.3. Effect of the adsorbent coverage on the heat of adsorption $-\Delta H$. Graphical presentation (present authors) of the data collected in Table 1 of Tewari et al. [37].

*IV.1.2. Uranium*

Uranium and associated radionuclides, in particular, radium and radon pose significant health risks to humans due to both radiotoxicity and chemical toxicity. Because isotopes of uranium have relatively long half-lives (ranging from approximately 4.5 billion years for $^{238}U$ to 245 thousand years for $^{234}U$ [38], human intake of uranium causes only slightly increased risk of cancer; however, a more immediate risk is posed by the potential for kidney damage resulting from chemical toxicity. The uranous form is highly insoluble, regardless of pH; however, oxidation to uranyl ($UO_2^{2+}$) results in moderate solubility [39]. Other chemical factors control the adsorption and desorption-and, therefore, mobility of uranium in water. Approximately neutral $pH$ and presence of minerals like $Fe$ oxides increase adsorption of $U^{4+}$, and decrease its mobility. In contrast, more alkaline conditions and increased bicarbonate concentration lead to greater desorption and mobility, especially of $U^{6+}$ [40].

The mixed oxide, magnetite ($Fe_3O_4$, $FeII/FeIII$) $= 0.5$ is assumed to be the main corrosion product of iron (considering the cooling systems of nuclear power plants) under anoxic conditions making its role in immobilizing uranium crucial. Because of the almost semiconductor character of magnetite surface, it potentially can function as mediator of electrons in the reduction on iron surface, leading to the precipitation of more insoluble $UO_2$ [41]. In addition, magnetite contains ferrous iron, which may also be able to reduce $U(VI)$ species. Rovira et al. [41] studied the interaction of $U(VI)$ in hydrogen carbonate medium with commercial magnetite as well as with magnetite formed as corrosion product on the surface of a steel coupon. The attention was paid to the effects of the hydrogen pressure and the mass of magnetite. The sorption of $U(VI)$ was performed at $25^oC$ in a solution containing $0.01 mM$ of $U(VI)$. The $pH$ of the solution ranged from $6$ to $7$ and the solution was continuously bubbled with $N_2$ or $H_2$. Parallel experiments, performed under $N_2 + CO_2$ or $H_2 + CO_2$ bubbling, showed that removal of uranium by magnetite was slightly more efficient under hydrogen only.(Fig. 4). The increase of hydrogen pressure in the system caused a faster decrease in the uranium concentrations in solutions. The higher pressures, the lower final uranium concentrations can be attained. The increase in the mass of the solid caused a slightly faster decrease of the concentrations of uranium in solution.

The source of carbonate seemed to have effect on the $U(VI)$ removal by magnetite. Similar tests, one

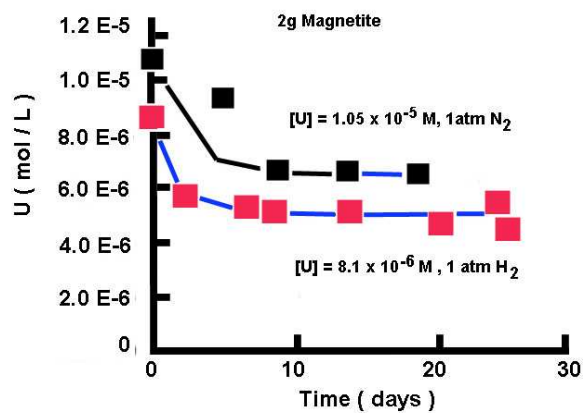

Fig.4. Uranium sorption onto magnetite nano-particles. Adapted from Rovira et al. [41]. Effect of $H_2(g)$ and $N_2(g)$ pressure on the concentration of uranium in $0.01M$ $NaCl$ solution ($\log pCO_2 = -3.82$)

continuously bubbled with $H_2 + CO_2$ gas mixture and the other with initial $NaHCO_3$ addition (but also bubbled with $H_2$), resulted in different evolution of the concentration of uranium in solution [41]. In sample with added $NaHCO_3$ the $CO_2$ was degassed by $H_2$ bubbling and the $pH$ was observed to increase. Therefore, the carbonate concentration decreased and the uranium concentration in solution decreased, too.

The interaction of $U(VI)$ solution with magnetite on the corroded steel coupon or with fresh steel coupon led to lower $U$ concentration in solution than in the case when only magnetite was used [41]. In the case of the magnetite on corroded steel surface, this was probably due to the presence of iron steel, which decreased the redox potential of the system [41],[42]. Rovira et al. [41] stated that the actual reductant in their experimental system was not hydrogen, but magnetite, ferrous iron provided by dissolution of magnetite, or iron in the case of having steel in the system. According to suggestion of Rovira et al. [41], when $U$ undergoes a reduction due to the presence of magnetite, the most likely precipitate is $UO_2(s)$.

Scott el al. [43] studied the interaction of $U(VI)$ with natural magnetite (the magnetite crystals were sectioned along 111, 110 and 100 crystal planes) as $3 mm$ thick coupons progressively cut and polished to 3-micron grade. Two sets of the micron grade coupons were exposed to the $20 mL$ of $0.042$ and $0.42 mM$ uranyl acetate solution at room temperature ($20^oC$) and $pH$ values $4.8$ and $4.2$, respectively. The experiments lasted $12$, $24$, $48$ and $168 h$. The X-ray Photoelectron Spectroscopy (XPS) and X-ray Diffraction (XRD) measurements indicated the presence of non-stoichiometric $UO_2$ at the surface of all the reacted magnetite coupons.

The experiments of Rovira et al. [44] indicate that the initial removal of uranium from solution is due to sorption onto the magnetite and then followed by reductive precipitation (supported by XPS and XRD data). As a stand of portion of the same series of experiments, El Aamrani et al. [45] investigated the interaction between $U(VI)$ and commercial magnetite in $0.1M$ $NaClO_4$ at $pH < 7$ in a recirculation reactor under $N_2$ atmosphere and the reaction lasted $57$ and $85$ days. In accordance with these experimental results the semi-conductive magnetite reduces $U(VI)$ to $U(IV)$ in the absence of metallic iron, but the removal of $U(VI)$ from solution is more efficient, when the metallic iron exists under the corrosion layer. Moreover, the reaction/removal rate increases, to some extent, when both the $H_2$ pressure and the mass of solids are increased. In this context, low carbonate concentrations augments the removal while increase of its amount led to hindering of the removal because the uranyl carbonate complexes stabilize the $U(VI)$.

Missana et al. [46] performed adsorption of $U(VI)$ on nanocrystalline magnetite ($50-200\,nm$) with emphasis on the adsorption kinetics and the kinetics of the $U$ reduction in the presence of magnetite. The sorption results indicate that the initial $pH$ of the solution affects the amount of uranium adsorbed at the beginning (1 day after the sorption starts). Precisely, at $pH = 7$ to adsorb about $90\%$ of the uranium only $24h$ are needed, whilst at $pH = 5$ within the same period of time this amount is only $40\%$. Missana et al. [46] observed about $80\%$ of the total amount of uranium adsorbed after 2 weeks, at both $pH$ mentioned above. According to these authors, the ionic strength of the solution on the uranium adsorption is negligible.

Leal and Yamaura [47] reported batch experiments ($30\,min$ of contact time) on adsorption of $UO_2^{2+}$ ions from nitric solution ($pH$ 4 and 5) onto magnetic nanoparticles with removal between 40% and 80%. The data are well described by equilibrium adsorption isotherm by either Freundlich or Langmuir models (see Fig.5a). The latter provides a maximum adsorption capacity of $27\,mg/g$. The effect of the contacting time is shown in Fig. 5b.

*IV.1.3. Cesium*

Catalette et al. [50] used natural magnetite with some impurities ($Fe_3O_4 : 96 \pm 0.6\ \%wt$, $SiO_2 : 2.4 \pm 0.5$, $CaO : 0.2 \pm 0.1$, $Al_2O_3 : 0.1 \pm 0.1$) in laboratory batch experiments for cesium sorption. The magnetite fraction $< 180\,\mu m$ exhibited about $18.3 \pm 0.2\,m^2/g$ specific surface area determined by the BET method. The point of zero charge $pH \approx 5.55 \pm 0.12$ was determined by potentiometric titrations at different ionic strength ($I = 10^{-2}, 10^{-1}, 1$). This value of PZC is low than $6.0$ commented earlier (see ref 6). The authors attributed it to the contamination of silica which even though its low mass percentage exhibits large specific area of about $400\,m^2/g$ [51] and low PZC about $pH_{pzc} = 2.0$.

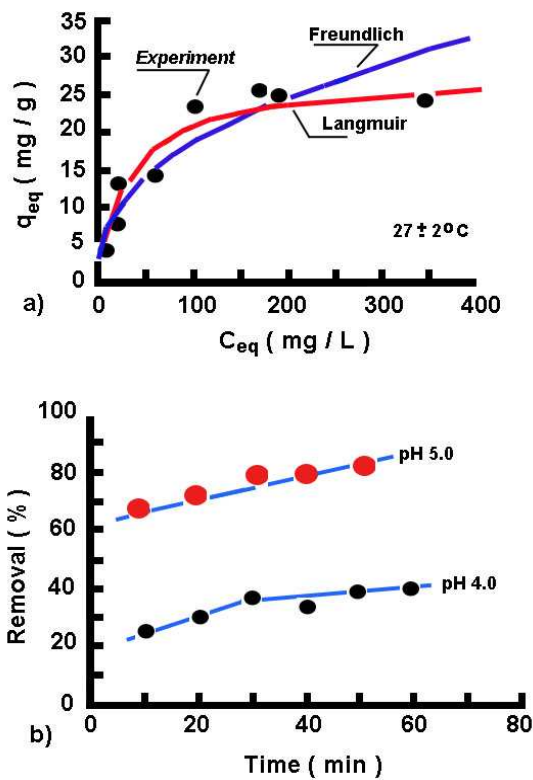

Fig.5. Uranium ($UO_2^{2+}$) $UO_2^{2+}$ removal by magnetite nano-particles. Adapted from Leal and Yamaura [47].
  a) Data fitting by two models of adsorption isotherms.
Effect of the contact time on the uranium uptake from nitric solution ($100\,mg/L$) at two $pH$ values.

Sorption of cesium was measured over the initial concentration range of. Since cesium is not sorbed at a ionic strength of $10^{-1}\,mol/L$, lower ionic strengths were used to minimize competition with other alkaline metal ions such as sodium. The results of Catalette et al. [50] are shown in Fig. 6a. According to these results, the more sodium decrease, the more cesium is sorbed on magnetite. The sorption is almost independent of the $pH$ which is not typical of a complexation site. The authors confirmed that they couldn't explain the results because the mechanism of this phenomenon was unclear. The silica effect was commented in the light of similar results reported by Todorovic et al., [52]. Moreover, to check the hypothesis of the silica effect, experiments with pure synthetic magnetite were performed with a much purer magnetite ($99.997\%$) with a specific area of about $2.0 \pm 0.2\,m^2/g$ ($BET\ N_2$) that resulted in an expected value of PZC confirming the literature data [51]. In the context of silica effect, experiments with admixtures of pure magnetite ($>99.92\%$, Merck, and surface area $380\pm10\,m^2/g$) and pure silica ($5\%wt$) resulted in cesium sorption quite close to that exhibited by the natural magnetite with impurities.

Marmier and Fromage [53] commented especially the cesium sorption on magnetite with silicate impurities. In a companion work [54] these authors reported that cesium does not bind to pure magnetite. However, they correctly mentioned that when impurities of silica exist like in the experiments of Todorovic et al. [52] and Catalette et al. [50], cesium sorption can be performed. The experiments performed reveal that silicates can bind on the magnetite surface over a wide $pH$ with a maximum sorption different from $100\%$. According to Marmier and Fromage [53], the greater affinity of cesium for silanol surface sites than for magnetite surface sites could explain the cesium sorption on natural magnetite with impurities (see ref.55). Similar to Catalette et al. [50] they performed a sorption experiments on an admixture of $300\,mg$ magnetite and $30\,mg$ silica and initial $Cs$ concentration $4\times10^{-5}\,mol/L$ (see Fig.6b) Experimental results show that the amount of cesium bound on the binary mixture of magnetite and silica is greater ($10$ to $20\%$ sorbed in addition) than the amount obtained by adding the contribution of the neat surfaces Fig.6c.

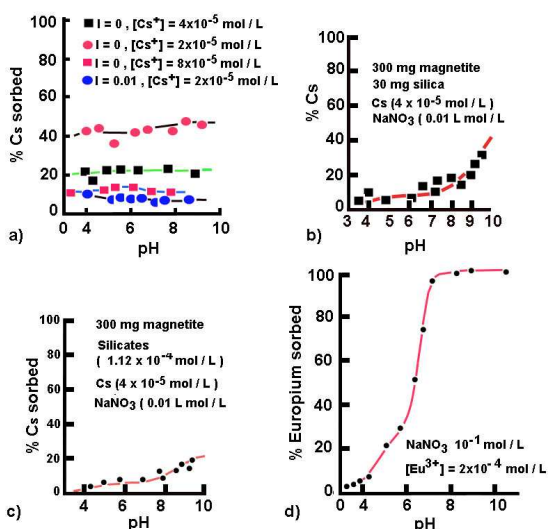

Fig. 6. Sorption of Cesium and Europium onto magnetite.
a) Sorption of $Cs^+$ onto $Fe_3O_4$ ($96\%$). Adapted from Catalette et al. [50].
b) Experimental and calculation results for the sorption of cesium. From Marmier and Fromage [53]. Solid label- experiments; Line- prediction by a surface complexation model
c) Sorption of cesium in presence of silicates. From Marmier and Fromage [53].
d) Sorption of Europium $Cs^+$ onto $Fe_3O_4$ ($96\%$). Adapted from Catalette et al. [50].

*IV.1.4. Europium*

Being a typical member of the lanthanide series, europium ($Eu$) usually assumes the oxidation state $+3$, but due to $Eu^{2+}$ ion electron configuration stability, europium also form common compounds in oxidation state $+2$, which all are slightly reducing. Europium has no significant biological role and appears not to be particularly toxic compared to other heavy metals and it is hard as lead, but is the most reactive of the rare earth elements and rapidly oxidize in air. Europium is produced by nuclear fission, too, but the fission product yields isotopes are low near the top of the mass range for fission products.

Catalette et al. [50] performed sorption experiments by using natural magnetite with silica impurities (see the information about the cesium sorption) for concentration ranging from $5\times10^{-5}\,mol/L$ to $4\times10^{-4}\,mol/L$ and two ionic strengths ($10^{-1}\,mol/L$ and $10^{-2}\,mol/L$). The sorption results are shown in Fig.6d for an initial europium concentration of $2\times10^{-4}\,mol/L$. The experiments of Hu et al. [56] on $Eu^{3+}$ adsorption onto silicates confirm the effect of the silica content in the experiments of Catalette et al. [50].

*V.1.5. Chromium*

Chromium is a member of the first row transition series of elements, which consists of Sc, Ti, V, Cr, Mn, Fe, Co, Ni, Cu and Zn, and belongs to group 6 of the periodic table, along with Mo and W. Chromium, which is one of the most toxic metals, is mixed into rivers and ground waters through the electroplating, metal finishing, leather tanning and chrome preparations. In the United States, it is the second most common inorganic contaminant in waters after lead [48]. Chromium is found in rocks, animals, plants, and soil. The most common forms of chromium are Cr (II), Cr (III), and Cr (VI), but exists mostly in two valence forms: $Cr(III)$ and $Cr(VI)$), out of which the later one is of the great concern due to its toxicity [49].

Namdeo and Bajpai [5] investigated deposition of a hexavalent chromium $Cr(VI)$ onto synthetic magnetite nano-particles from aqueous solutions in the range $30-50^oC$. The adsorption process follows Langmuir-type behavior (see Fig.7a) and the adsorption capacities $1.526$, $3.069$ and $3.957\,mg/g$ for the runs performed at $30$, $40$ and $50^oC$, respectively. The intraparticle diffusion of $Cr(VI)$ was confirmed by linear nature of Bangham (see Fig.7b,c). The sorption free energy calculated by Dubinin-Radushkevich is $13.51\,kJ/mol$ and indicates the chemical nature of the adsorption process.

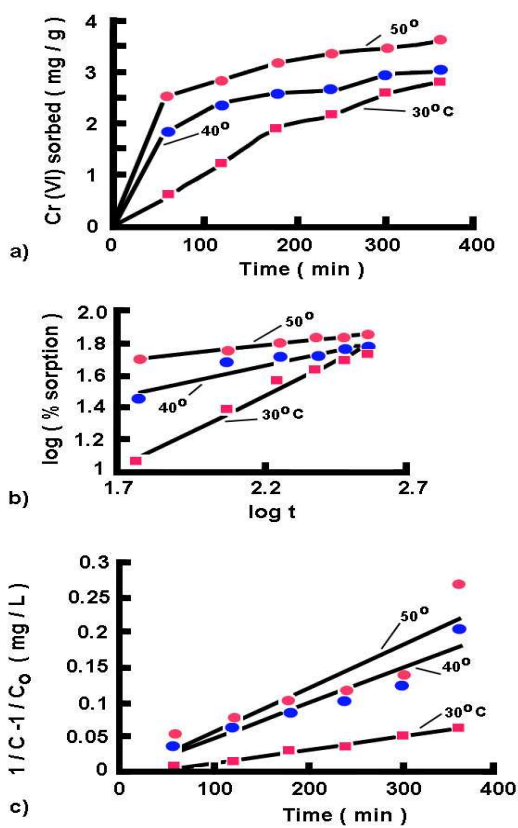

Fig. 7. Chromium uptake from aqueous solution by magnetite. Adapted from [5]. Effect of temperature.

Amin et al. [57] reported removal of $Cr(VI)$ by magnetite nanoparticles ($40-300\,nm$) from simulated electroplating wastewater (about $10\,mg/L$). Only $1\,g/L$ magnetite particles were enough to remove about $82\%\ Cr(VI)$. According to these authors Competition from common coexisting ions such as $Na^+$, $Ni^{2+}$, $Cu^{2+}$, $NO_3^-$, $SO_4^-$, and $Cl^-$ negligible. Moreover, the rate of chromium removal is reduced as the $pH$ in the solution increases. The Freundlich isotherms used to describe the equilibrium provided parameters depending on the value of $pH$ (see Fig. 8). As pointed out by Amin et al. [57] in highly acidic media, the adsorbent surfaces can be highly protonated thus able to uptake Cr(VI) in the anionic form, $HCrO_4^-$ [58]. Hence, with increase in $pH$ the magnetite surface becomes more positively charged and the attraction of negatively charged $Cr(VI)$ anions will be enhanced [59].

### IV.1.6. Selenium

Selenium is a metalloid found in Group VIA of the periodic table, below sulfur. It has similar chemical properties to sulfur due to its analogous electron distribution, which can make remediation in the presence of sulfur challenging. Selenium is present in the environment in both inorganic and organic forms, and in the solid, liquid, and gas phases. Selenium is a natural trace element found in bedrock, but it is also introduced into the environment by anthropogenic activities, such as mining and combustion of fossil fuels [60]. The environmental impact of Se in a given environment strongly depends on its speciation and concentration. Because Se is an essential element, organic Se compounds can be found in both living and dead organisms [61] and in dissolved phases [62]. In decomposition processes, $Se^{2-}$ is transformed into $Se(IV)$ ($SeO_3^{2-}$) or $Se(VI)$ ($SeO_4^{2-}$). The former one, is more strongly adsorbed on the surface of iron oxyhydroxides and the adsorption process is largely $pH$. Both selenate $Se(VI)$ and selenite $Se(IV)$ are bioavailable and have high potential for bioaccumulation and toxicity.

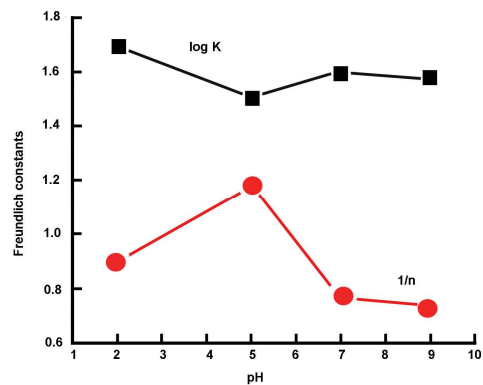

Fig. 8. The Freundlich constants at different $pH$ in sorption of $Cr(VI)$ onto magnetite. Experiments of Amin et al. [57. Graphical presentation of data summarized in Table 3 of [57].

Martinez et al. [63] studied the sorption of $^{79}Se$ ($Se(IV)$ and $Se(VI)$ onto magnetite. Magnetite is a mineral present in the near-field of a nuclear waste repository that might represent an important retardation factor for the mobility of many radionuclides. For example $^{79}Se$ is one of the principle component of the radioactive wastes and the main selenium radioactivity will be due to the $^{79}Se$ isotope (with a half-life of $6.5x10^4$ years). Two main issues were considered by Martinez et al. [63], namely:

i) the effect of the initial selenium concentration in solution, and
ii) the effect of pH on the sorption process.

Pure magnetite with particle size $<5\mu m$ and surface area (BET) of about $0.89\pm0.01\,m^2/g$ was tested for sorption (batch experiments at room temperature) from model solutions prepared dissolving either $Na_2SeO_4\cdot 10H_2O$ or $Na_2SeO_3$ in *Milli-Q* water. The $Fe(II)/Fe(III)$ ratio measured in the surface of the solids by using XPS indicated the theoretical value for pure magnetite. The equilibrium was reached at least for $30h$ ( see Fig.9a). The $pH$ of the solution was varied (between $2$ and $12$) by adding $HCl$ or $NaOH$ but the ionic strength was maintained at $0.1 mol/dm^3$ $NaCl$ (Fig.9b). The sorption equilibria were successfully fitted by the Langmuir isotherm (See Fig. 9c) based on the following equilibrium

$$Se + S \Leftrightarrow S - Se \qquad (20)$$

$$K_L = \frac{\{S-Se\}}{\{S\}\{Se\}} \qquad (21)$$

where $\{S-Se\}$ stands for the concentration of the occupied surface sites, while $\{S\}$ is for the free surface sites. $K_L$ is the Langmuir constant ($dm^3/mol$).

According to Martinez et al. [63], the Langmuir constant, $K_L$ for magnetite is higher that those obtained with sorption of $Se(IV)$ and $Se(VI)$ onto hematite and goethite [64]. In this context, for instance it as established that for the Langmuir isotherm modelling of the sorption parameters are:

$\Theta_{max} = 3.13 \pm 0.07 \times 10^{-7}$ $mol/m^2$ and
$K_L = 1.19 \pm 0.07 \times 10^6$ $dm^3/mol$ for $Se(IV)$;

$\Theta_{max} = 3.5 \pm 0.02 \times 10^{-7}$ $mol/m^2$ and
$K_L = 3 \pm 0.01 \times 10^5$ $dm^3/mol$ for $Se(VI)$.

Jordan et al. [65] reported $Se(IV)$ adsorption by magnetite (commercially provide powder with surface area $1.6 m^2/g$) in presence of silicic acid under the solubility limit of amorphous silica. The main outcomes of this study are:

- Experimental results were modeled by using surface complexation models (SCMs), which consider functional surface groups as complexing ligands in solution.
- With $pH = 4.0$ the sorption of $Se(IV)$ onto magnetite reaches a saturation after $30h$.
- The sorption onto magnetite decreases with increase in $pH$ of the solution as it was expected for anions (Martinez et al, 2006) - see Fig. 10.
- The silicic acid can bind on the surface of magnetite over a wide $pH$ range. This differs from the classical adsorption where the sorption classically increases or decreases in a sharper $pH$. Maximal percentage of silicate sorption was observed in the range $pH$ $8$ - $10$, thus including the first $pK_a = 9.86$ of $H_4SiO_4$ [66].
- Owing to reducing properties of magnetite, $Se(IV)$ could be reduced into elementary selenium during sorption experiments but the XPS study indicated no reduction for a three month equilibration time.

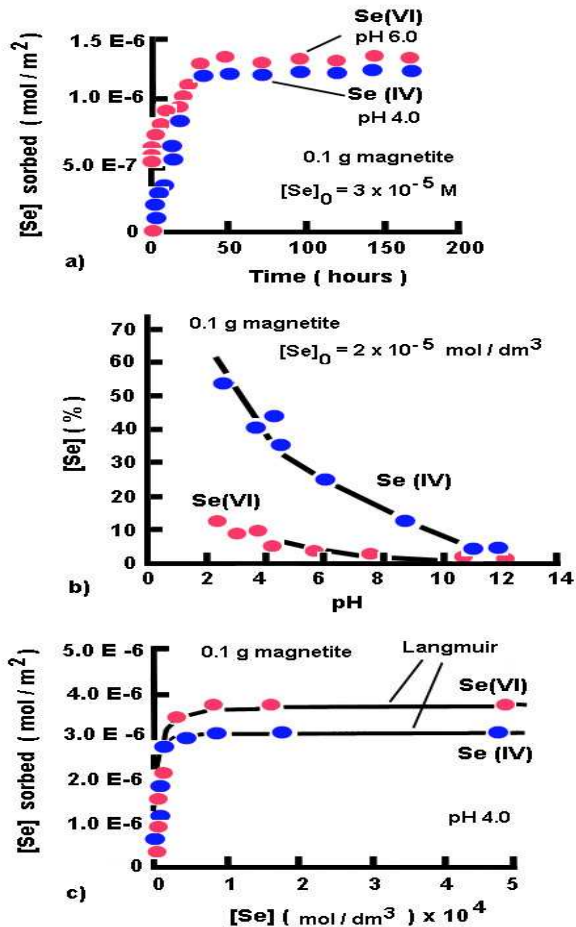

Fig. 9. Selenium sorption of Selenium onto magnetite. Adapted from Martinez et al. [63]
    a) Time evolution of selenium sorbed.
    b) Sorption of selenium onto magnetite a function of $pH$
    c) Sorption isotherms for selenium onto magnetite.

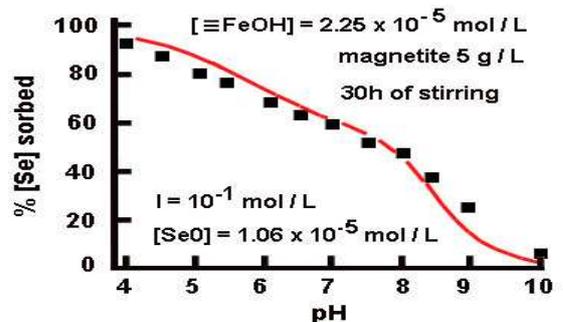

Fig. 10. Sorption of selenium(IV) onto magnetite as a function of the $pH$. Adapted from Jordan et al. [66].

*IV.2. Arsenite and Arsenate Removal*

Arsenic is a heavy metal occuring in the environment in different oxidation states and species, e.g., $As$ as $As(V)$, $As(III)$, $As(0)$ and $As(-III)$ ). Moreover, arsenic cannot be easily destroyed but can only be converted into different forms or at least transformed into insoluble compounds by contacts with other elements, such as iron, for example. Inorganic arsenic generally exists in two predominant oxidation states, **arsenite** $NaAsO_2$ and **arsenate** $Na2HAsO_4$, both of which are toxic to man and plants [67]. Arsenic is a metal that can generate multiple adverse health effects because of the many chemical forms it takes on. For example, arsenic can appear in inorganic or organic form. In oxidized environment $As$ appears mostly as oxyanions [68]. Arsenic trioxide, sodium arsenite, and arsenic trichloride are the most common inorganic trivalent arsenic compounds. Trivalent compounds of arsenic are the most toxic forms of arsenic. Inorganic arsenic is always considered a potent human carcinogen, associated with increased risk for cancer of the skin, lungs, urinary bladder, liver and kidney [67]. Arsenic-contaminated groundwater, used as drinking water, has been a severe problem worldwide, especially in Bangladesh, India, some parts of Europe, South America and United States [67],[69]. Conventionally, there are several methods for arsenic removal [67], among them. These methods include coagulation and flocculation, precipitation, adsorption and ion exchange, membrane filtration. In the context of the topic of the present analysis, we address on adsorption, mainly by iron oxides, and onto magnetite in particular, among these sorbents.

*IV.2.1. Adsorption*

The iron oxides have a high affinity for the adsorption of arsenite and arsenate [70], [71]. Yean et al. [72] studied the effect of magnetite nanoparticles size on the adsorption and desorption of arsenite and arsenate for water cleaning purposes. It was observed that decrease in the particle size results in increased adsorption maximum capacity for both arsenite and arsenate. On the other hand, the arsenic desorption was hysteretic an effect demonstrated strongly with decrease the particles in size. According to these authors, the hysteresis is due to higher arsenic affinity of the magnetite nanoparticles. All adsorption data were fitted with the Langmuir isotherm (Eq.22) (as it is shown in Fg.11a,b ), namely

$$q = \frac{bq_{max}C}{1+bC} \qquad (22)$$

The $As(III)$ adsorption demonstrated pH independence and the equilibria were modeled by the Langmuir adsorption isotherm only. The maximum adsorbed amount on the $300\,nm$ particles was about 18 time lower that that on the $20\,nm$ samples- see Fig.11c,d. At the same time the surface areas of both factors differed by factor of 16 that simply explains the gross adsorption performance. The pH independency of $As(III)$ was attributed to the high $pKa$ value $As(III)$, i.e. $pK_{a,1} \approx 9.22$ that indicate the arsenite mostly in neutral specie up to $pH \approx 8.0$. The latter indicate that this non-ionic form of $As(III)$ does not affect the $pH$ of the solution. However, with adsorption of $As(V)$ the maxima in the adsorption capacities of both magnetite fractions decrease as the $pH$ of the solution is increased. The magnetite surface is positively charged up to $pH \approx 6.8$ due to the **point of zero charge (PZC)** of the magnetite [5],[6]. Therefore, decreasing the $pH$ more $As(V)$ can be adsorbed. At higher $pH$, more precisely beyond $pH \approx 8.0$, the repulsion of the negatively charged $As(V)$ from the negative sites at the magnetite surface reduces the adsorbed amount. Both, $As(III)$ and $As(V)$ demonstrated stronger affinities to adsorb to the fine fraction ($20\,nm$) demonstration larger values of the $b.q_{max}$ product (see the Langmuir isotherm- eq. 22) in the range from 0.23 to 2.09 $L/m^2$. Internal, at each $pH$ value tested, the weight based ($As(III)$ and $As(V)$) absorption capacities ($q_{max}, \mu mol/g$) were larger and attributed to the higher surface area of the finer fraction. On the other hand, the surface-area based adsorption capacities ($q_{max}, \mu mol/m^2$) are almost similar for both fractions ($20\,nm$ and the $300\,nm$) but with a tendency to decrease with increasing $pH$.

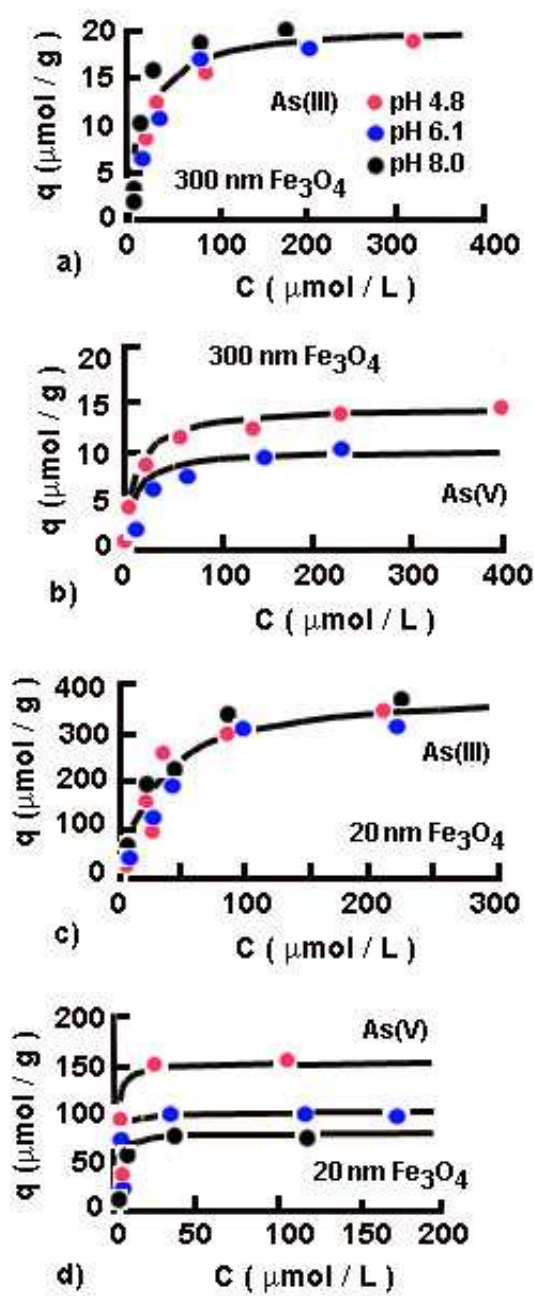

Fig.11. Arsenite and arsenate adsorption onto magnetite. Adapted from Yean et al. [72].
The solid curves are drawn using the curve-fitted Langmuir model.

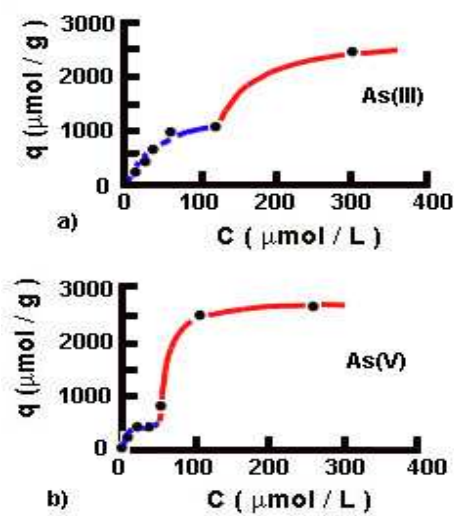

Fig.12. Plot of the adsorption of $As(III)$ and $As(V)$ onto $11.72\,nm$ magnetite at $pH$ 8.0 H. Adapted from Yean et al. [72]. The sections of the two-staged isotherms are presented in different colours.

In contrast, to the results just analyzed, the use of finer particles ($11.72\,nm$) led to different adsorption isotherms shown in Fig.12a,b. Yean et al (2005) described them as sums of two Langmuir isotherms. The first (the lower) isotherm (**blue colour**) describes well the adsorption equilibria at low concentrations: $<120\,\mu mol/L$ for $As(III)$ and $<50\,\mu mol/L$ for $As(V)$. Within this range, the maxima in the adsorption capacities were $1532\,\mu mol/g$ and $622.7\,\mu mol/g$ for $As(III)$ and $As(V)$, respectively. The latter are equivalent to 15.5 and 6.3 $\mu mol/m^2$, respectively. The second isotherm (**red colour**) fitted quite well the data beyond $>120\,\mu mol/L$ for $As(III)$ and $>50\,\mu mol/L$ for $As(V)$. This adsorption tests within concentration range resulted in substantially higher maxima in the adsorption capacities and was attributed to co-precipitation of $As$ with $Fe$ on the surface [73], [74].

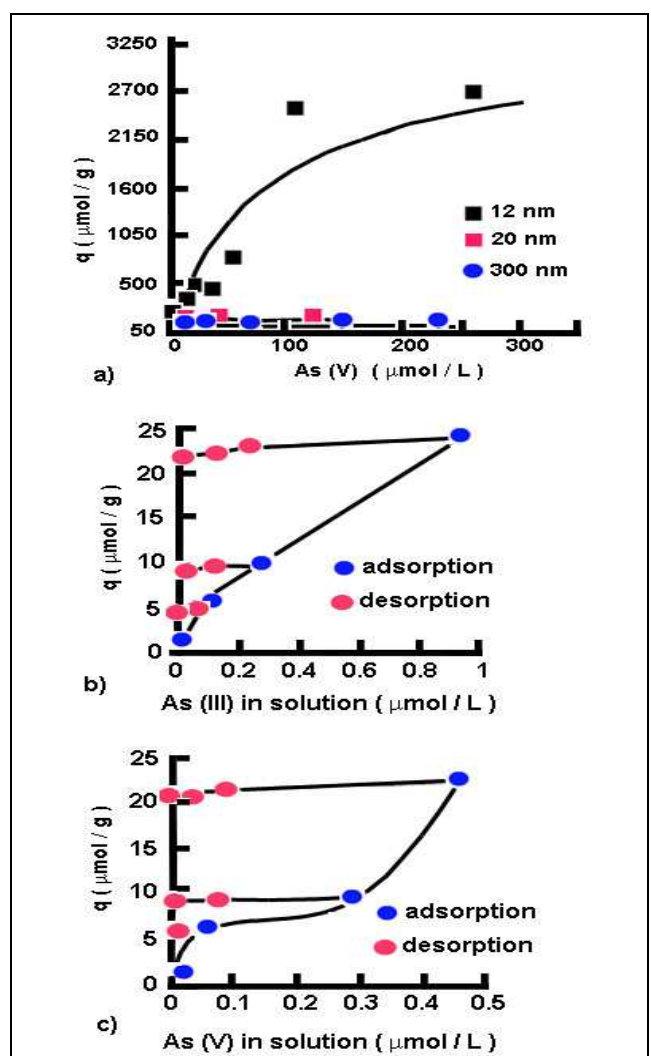

Fig.13. Adsorption of $As(III)$ and $As(V)$ onto microcrystalline magnetite. Adapted from mayo et al. [75].
a) As(V) adsorption on different magnetite nanoparticles. The solids lines are the adsorption isotherms modeled by the Langmuir equation.
b) Adsorption and desorption of $As(III)$ onto $20\,nm$ magnetite.
c) Adsorption and desorption of $As(V)$ onto $20\,nm$ magnetite.
Note: all data are those corresponding to the equilibrium, i.e. equilibrium adsorbed vs. the equilibrium in the solution.

The magnetite nano-cristalinity on the arsenic adsorption was studies by Mayo et al. [75] in solutions with $As$ concentration ranging from $0$ to $0.45\, mmol/L$ and solids $2.5$ and $0.1\, g/L$ for commercially made $300\, nm$ and $20\, nm$ $Fe_3O_4$, respectively. All adsorption isotherm data were fitted by the Langmuir isotherm equation (see Fig.13a). In general, an increase in the weight-based $As(V)$ adsorption density was observed with decreasing in the particle size. This fact was attributed to the larger surface area exhibited by the finer particles.

*IV.2.2. Desorption*

Desorption experiments [72] with the larger fractions of magnetite nanoparticles ($20\, nm$ and $300\, nm$) exhibited hysteretic behaviour. With the larger fraction ($300\, nm$) after three steps of desorption, each taking $24h$, only $20-25\%$ of the adsorbed $As(III)$ or $As(V)$ was desorbed. Moreover, with the fine fraction ($20\, nm$) for both of both $As(III)$ and $As(V)$ a complete adsorption hysteresis was observed. The latter implies, that after three steps (each of 24 h), only about 1% of the adsorbed $As(III)$ or $As(V)$ was desorbed.

The desorption hysteresis reveal that the finer magnetite particles exhibit higher adsorption affinity to both $As(III)$ and $As(V)$. According to Yean et al. [72] the $20\, nm$ fraction has advantage with respect to that consisted of $300\, nm$ in two aspects: stronger adsorption due to large contact area and practically negligible adsorption. The latter could be considered as a useful characteristic in the waste disposal.

Similar results was observed by Mayo et al. [75], i.e. almost irreversible adsorption with about 1% desorbed $As$ at $pH\, 6.1$ as well as when $pH$ was changed to $pH\, 4.8$ and $pH\, 8.0$ (see Fig.13b,c). This led to the conclusion that such particles can be used for water treatment.

## V. Separations

*V.1 Some general comments on sorption process*

The separation of magnetic particle from the aqueous solutions and safe deposition are not, in general commented by the articles reviewed in the preceding sections. To some extent, some authors mention about consequent high-gradient magnetic separations commonly at a laboratory scale. Therefore, as a natural consequence of the analysis of the absorption processes we like to stress the attention on three principle issues affecting both the adsorption process performance and the separation step concerning magnetic particle removal from the aqueous solutions, namely :

- Adsorption process itself as
- Separation from the solution and its link to preceding adsorption step
- Alternatives in sorption process organization

Adsorption processes employing magnetite for sorption of contaminants from wastewaters work principally in two operating modes [76]:

a) **Perfect mixing in tanks** (batch operations) followed by separations either mechanical or magnetic. This operating mode needs huge volume for the adsorption stage and consequent organization of the separation process either by filtration (membranes) or magnetic methods [77]. We limit our comments only to these notes and the readers interested in large scale separations by iron oxides and particularly by magnetite can find enough information elsewhere[78], [79]. Some comments on HGMS are given next

- **HGMS** Conventional magnetic separation processes use, for example, fine stainless steel wool to form a magnetic matrix within a flow field of a solution containing mineral particles to be separated. In high-gradient magnetic separation (HGMS) the particle collection efficiency depends on the matrix parameters such as magnetic material, polymer mesh pattern, matrix lamination pattern; fluid speed and sizes and external magnetic field (see [80]-[87]). Magnetic separation is a powerful method utilized from long time in the treatment of strongly magnetic mineral ores and for the removal of ferromagnetic impurities from mixtures.

The conventional HGMS processes use a fine steel wool or affixed bed of steel spheres to form a magnetic matrix to create a deep-bed filter with strong magnetic gradients in the pore space where the contaminated fluid flows. A common steps in this process is the preliminarily flocculation in order to crate large particles ($>1\mu m$ in diameter). This approach is now widely applicable when magnetite nano-particles are used as adsorbents. With coarse magnetic adsorbents, either HGMS is the suitable separation techniques if large amounts of water have to be treated, for instance, or other techniques from the conventional "arsenal" of the chemical engineering can be applied.

b) **Adsorption in fixed bed** operations (in columns) where the deposited substances remain on the magnetic grain surfaces and can be safely deposited after that. With respect to the magnetic adsorbent this operation is batch one, while the wastewater flow continuously through the bed up the break-through point [88],[89],[90],[91]. For column operations, crystalline ferrites and powdered magnetite must be supported with another material to achieve pressure drop and good water flow through the column. The first experiments of supported magnetite involved using a column containing a nonporous polyamine-epichlorohydrin resin bead coated with activated magnetite surrounded by an electromagnet of $0.3T$ [92]. The work of Cotten et al.[32] is a such example where $50\%$ of the bed are glass beads.

In the context of the previous comments, it is noteworthy to stress the attention that the classical fluid-solids contacting techniques using fixed or fluidized bed operations are limited flow bellow by the size of particles used. Decreasing the particle seize we get high surface area required for efficient sorption process but at the same time this leads to reduction in the fluid flow rate which is limited from above by the minimum fluidization velocity. Moreover, the decrease of particles in size results in bad flow distribution trough the bed cross-section, channeling and fluid bypass. The sorption process should avoid the solids backmixing, so any problems emerging in the organization of good fixed bed operation should be avoided.

*V.II. Coarse vs. nano-particles*

Therefore, the main problem in the sorption by nano-particles is the impossibility to organize efficient fixed bed operation because this contacting technique works quite well with particle sizes beyond $100 \mu m$. Even with some micron-sized natural magnetite particles, the channeling and the limited flow rate range do not allow the fixed bed operation to be applied.

Hence, from the point of view of process organization, the nanoparticles can operate in the batch perfect mixing mode, that is in fact, the same as in the laboratory flask, but a larger scale, and then removed by HGMS. The HGMS is not low-cost equipment and its application should carefully analyzed before designing the entire separation process. On the other hand, the coarse particles works quite well in fixed beds that is proved by many sorption operations.

$Fe_3O_4$ (magnetite) particles having a size of a few micrometers is typical for the crushed rocks and "black beach" native magnetite deposits were tested as adsorbents for the removal of radionuclides and arsenic from contaminated water [93]. Albeit their efficient use the adsorption separation processes [72] [94] the reduction in size (increasing the solids contact area) calls for new magnetic recovery techniques. From the data analyzed in this review, it becomes evident that any impurities by carbonates, silicates, etc. on the surface of the natural magnetite particles augment the sorption process in contrast to the pure synthetic adsorbents. Therefore, what has to be done: To use natural magnetite with impurities and organize efficient fixed bed adsorption process with high throughput and efficiency or to use fashionable nano-scale magnetite that does not allow escaping the scale of the laboratory flask and requiring HGMS equipment? What has to be the technical solution following the laboratory experiments? Nowadays, the answer is straightforward: the nanoparticles are better due to the high surface area provided but inapplicable in the large scale processes. The coarse particles are natural, have some impurities and work well in fixed beds, despite the fact that their specific surface area is lower that exhibited by the nano-scaled counterparts. As usual, in the modern world, the common answer is: the choice is yours! However, the realistic engineering should take into account all issues mentioned above and take a realistic decision.

## VI. An alternative in adsorption-separation technique

The alternative approach in the sorption with magnetic adsorbents should compromise the fixed bed operation properties avoiding the solids backmixing and working with coarse particles, and the compact mass of particles required by the magnetic recovery. In the context of this statement we address the magnetic field assisted fluidization [95],[96],[97]. This technique allows a bed of magnetic particles to operate under fixed bed condition beyond the point of minimum fluidization in absence of field. The necessary tool to attain such desired conditions is the application of an external magnetic field (either axial or transverse) [95],[96],[97]. The high flow rate possible in the so-called magnetically stabilized beds (MSB) allows compensating the "deficiency" due to the coarse particles used instead nano-scaled counterparts. This simply means, that the high fluid-solid slip velocity allows to diminish the external mass transfer resistance (diffusion in the liquid phase) and to work almost under a kinetic control of the sorption process. In contrast, with nano-particles, the stronger mass transfer resistance is near the fluid-solid surface were the stagnant liquid is almost saturated.

The MSB operation practically avoid the used the of the magnetic recovery of fine particles from large liquid volumes, a practice discussed at large in this review and commonly used in wastewater treatment by iron oxide as sorbents. The particle bed is in a column and the waste flow just pass through it at high flow rate. When the adsorption process is over with a given bed attaining saturation just as witching of the flow to another column is necessary. Because, in the case of hazardous contaminants, such those discussed here, the desorption and sorbent recovery are undesirable, just a remote removal of the saturated bed and a consequent safe deposition is necessary. The magnetically assisted fluidization techniques require much lower energy than HGMS ad use coarse particles. Hence, recall about the natural magnetite and the other oxides, discussed in this work, which leads to the idea of low-coast and high-efficient operations for decontamination of large wastewater volumes. Separations of various species from either gaseous or liquid flows are well analyzed [97], [98].

## VII. Conclusions

The review performed went through a very narrow area of separation (adsorption) technique considering magnetite only as adsorbent. The choice of the adsorbent material was special because magnetite can be obtained either naturally or through a synthesis. The data analyzed indicate that the old-fashioned natural magnetite with some impurities can adsorb better than the pure synthetic nano-particles. However, the studies on the adsorption on nano-particles dominate in the literature. Moreover, these studies are limited only to sorption experiments in small volumes and despite the excellent results the consequent step towards large scale applications is still unclear.

The pure adsorption studies indicate that the most used equilibrium model is the Langmuir isotherm and the Freundlich one, to some extent. In this context, a deep analysis on the adsorption processes at this scale of the adsorbent and various hazardous contaminants which will allow collating many data from a unified point of view is still missing.

The second reason to address only magnetite particles as sorbents was the suggested here idea for alternative separation process in magnetically stabilized bed. Beds of magnetite are well studied and work very well under magnetic field assisted fluidization [95]-[98].

## References

[1] U. Schwertmann, R.M. Cornell, *Iron Oxides in the Laboratory: Preparation and Characterization*," VCH Publishers, Inc., New York (1991).
[2] Z. Sun, F. Su, W. Forsling, P. Samskog, Surface Characteristics of Magnetite in Aqueous Suspension, *Journal of Colloid and Interface Science, 197* (1998)151 -159.
[3] P.J.Vikesland; A.M. Heathcock; R.L.Rebodos; K.E. Makus, Particle size and aggregation effects on magnetite reactivity toward carbon tetrachloride, *Environmental Science and Technology 41* (2007) 5277-5283.
[4] M.L.Peterson; A.F.White; G.E.Brown, G.A.Parks, Surface passivation of magnetite by reaction with aqueous Cr(VI): XAFS and TEM results, *Environmental Science and Technology, 31* (1997) 1573-1576.
[5] M. Namdeo,S.K. Bajpai , Investigation of hexavalent chromium uptake by synthetic magnetite nanoparticles, EJEAFChe, 8 (2009 ) 367-381.
[6] R. M. Cornell, U. Schwertmann, *The Iron Oxides, Structure, Properties, Reactions, Occurrences and Uses*, Wiley-VCH, 2nd Edition, 2003.
[7] L. Babes, B Denizot, G. Tanguy, J J. Le Jeune, P. Jallet, Synthesis of iron oxide Nanoparticles used as MRI contrast agents; Aparametric study, *Journal of Colloid and Interface Science* 212 (1999) 474-482.
[8] A. A. Olowe , J. M. R. Genin, The mechanism of oxidation of ferrous hydroxide in sulphated aqueous media: important of the initial ratio of the reactants, *Corrosion Science, 32* (1991) 965-984.
[9] A. E. Ragazzoni, G. A. Urrutia, M. A. Blesa A. J. G. Maroto, Some observations on the composition and morphology of synthetic magnetites obtained by different routes, *Journal of Inorganic Nuclear Chemistry, 43* (1981) 1489-1493.
[10] S. L. Brantley, L. J. Liermann, R. L. Guynn, A. Anbar, G. A. Icopini and J. Barling, Fe isotope fractionation during mineral dissolusion with and without bacteria, Geochemica et Cosmochimica Acta, 68(2004) 3189-3204.
[11] Stumm, W., J. J. Morgan, Aquatic *Chemistry: Chemical Equilibria and Rates in Natural Waters*, Wiley-Interscience publications, 1996.
[12] N. Das, N., R. K. Jana, Adsorption of some bivalent heavy metal ions from aqueous solutions by manganese nodule leached residues. Journal of Colloid and Interface Science **293**(2006) 253-262.
[13] Lagergren, S. (1898). Zur theorie der sogenannten adsorption gelöster stoffe. Kungliga Svenska Vetenskapsakademiens, *Handlingar 24* (1898) 40.
[14] M. Kosmulski, *Chemical properties of material surfaces*, Marcel Dekker, New York, 2001.
[15] U. Schwertmann , Iron Oxides, in: Chesworth W., Encyclopedia of Soil Science, Springer, Netherland, (2008) pp. 363-369.
[16] J.J. Lyklema, A. De Keizer, B.H. Bijsterbosch, G.J. Fleer, M.A. Cohen Stuart, Electrokinetics and related phenomena, In: Fundamentals of Interface and Colloid Science (*FICS*). (1995), 413-414. Academic Press .
[17] E. Tombácz, E. Illés, A.Majzik, A. Hajdú, N. Rideg, M. Szekeres , Ageing in the Inorganic Nanoworld: Example of Magnetite Nanoparticles in Aqueous Medium , *Croatica Chemica Acta , 80* (2007) 503-515.
[18] M. Kosmulski, pH-dependent surface charging and points of zero charge. IV. Update and new approach, *Journal of Colloid and Interface Science, 337* (2009) 439–448. doi:10.1016/j.jcis.2009.04.072.
[19] C. W. Turner, D. H. Lister, D. W. Smith; The Deposition and Removal of Sub-Micron Particles of Magnetite at the Surface of Alloy-800 , Steam Generator and Heat Exchanger Conference, Toronto, Canada, 1990.
[20] T. M. Riddick; Control of Colloid Stability through Zeta Potential ", Livingston Publishing Company, 1968.
[21] E. Tombácz, Adsorption from Electrolyte Solutions, Ch.12. In: J. Tóth (Ed.) *Adsorption: Theory,Modeling, and Analysis*, Marcel Dekker, New York, 2002. pp. 711–742.
[22] P. Tartaj, M.P.Morales, S. Veintemillas-Verdaguer, T. Teresita Gonzalez-Carreno, J. Carlos Serna, The preparation of magnetic nanoparticles for applications in biomedicine, *J. Phys. D: Appl. Phys. 36* (2003**)** R182–R197.
[23] S.Y.Shaw, Y.J.Chen, J.J. Ou, J.J., J.Ho, Preparation and characterization of *Pseudomonas putida esterase* immobilized on magnetic nanoparticles, Enzyme Microb. Technol. , 39(2006) 1089–1095.
[24] J. Lee, T. Isobe, M. Senna (1996) Magnetic properties of ultrafine magnetite particles and their slurries prepared via in-situ precipitation , *Colloids and Surfaces A: Physicochemical and Engineering Aspects, 109* (1996) 121-127
[25] C. Yu, X. Hao , H. Jiang , L. Wang (2011) $Fe_3O_4$ nano-whiskers by ultrasonic-aided reduction in concentrated NaOH Solution , Particuology 9 (2011) 86–90. doi:10.1016/j.partic.2010.06.003.
[26] T.K. Indira, P.K. Lakshmi , Magnetic Nanoparticles-A Review ,International Journal of Pharmaceutical Sciences and Nanotechnology , 3 (2010) 1035-1042.
[27] Q. Yuan, R. A. Williams , Large scale manufacture of magnetic polymer particles using membranes and microfluidic devices , *China Particuology 5* (2007) 26-42.
[29] Y.F. Shen, J.Tang, Z.H. Nie, Y.D.Wang, Y. Renc, L. Zuo, Preparation and application of magnetic Fe3O4 nanoparticles for wastewater purification, *Sep. Purif. Technol., 68* (2009) 312–319.
[30] S. Conroy, S.H.Jerry Lee, M. Zhang, Magnetic nanoparticles in MR imaging and drug delivery. *Adv. Drug Deliv. Rev., 60* (2008) 1252–1265.

[31] J.D. NAVRATIL, Removal of Impurities Using Ferrites and Magnetite, Australian Patent Application PJ0198 (August **8,** 1988)

[32] G. B. Cotten , J. D. Navratil, H. Bradley, Magnetic Adsorption Method for the Treatment of Metal contaminated aqueous waste , Waste Management '99 Conference, March 1999, Tucson, Arizona,

[33] Petkovic D.M., S.K. Milonic, Adsorption of cesium from basic water solutions on natural magnetite, *Bulletin of Boris Kidric Institute of Nuclear Sciences, 20* (1969) 17-23.

[34] S.K. Milonic, A. Lj. Ruvarac, Adsorption of $Cs^+$, $Co^{2+}$ and $Ce^{2+}$ from acid aqueous solutions on natural magnetite, *Bulletin of Boris Kidric Institute of Nuclear Sciences, 21* (1970) 21-26.

[35] S.K. Milonic, M.M. Kopecni,(1983) The point of zero charge and adsorption properties of natural magnetite, *J Radioanalytical Chemistry, 78* (1983) 15-24.

[36] H. Tamura, E. Matijevic, L. Meitess, Adsorption of $Co^{2+}$ ions on spherical magnetite particles , *J. Colloid and Interface Sci. 92* (1983) 303-314.

[37] P. H. Tewari, A. B. Campbell, W. Lee, Adsorption of $Co^{2+}$ by Oxides from Aqueous Solution, *Canadian Journal of Chemistry, 50* **(**1972) 1642-1648.

[38] S.A.Copenhaver, S. Krishnaswami, K.K.Turekian, N. Epler, J.K. Cochran, Retardation of 238U and 232Th decay chain radionuclides in Long Island and Connecticut aquifers, *Geochimica et Cosmochimica Acta, 57*(1993) 597-603.

[39] D. Banks, O.Royset, T.Strand, H.Skarphagen, Radioelement (U, Th, Rn) concentrations in Norweigan bedrock groundwaters, *Environmental Geology, 25* (1995) 165-180.

[40] R.B. Wanty, S.L.Johnson, P.H.Briggs, Radon-222 and its parent radionuclides in groundwater from two study areas in New Jersey and Maryland, U.S.A.: *Applied Geochemistry, 6* (1991) 305-318.

[41] M.Rovira, J.de Pablo, S. El Armani, L. Duro, M. Grive, J.Bruno, Study of the role of magnetite in the immobilization by reduction to U(VI) under the presence of H2(g) in hydrogen carbonate medium. SKB Technical Report TR-03-04, January 2003, Swedish Nuclear Fuel and Waste Management Co (SKB), 2003, Stockholm, Sweden.

[42] E. Myllykylä , Reduction of Uranium in Disposal Conditions of Spent Nuclear Fuel , Working Report 2008-09, February , 2008 , VTT Technical Research Centre of Finland , Fosiva OY, Olkiluoto, Finland .

[43] T.B. Scott, G.C.Allen, P.J. Heard, M.G. Randell,Reduction of U(VI) to U(IV) on the surface of magnetite. *Geochimica et Cosmochimica Acta, 69* (2005),5639–5646.

[44] M.Rovira, S. El Armani, L. Duro, L.Gimenez, J. de Pablo, J. Bruno, Interaction of uranium with in situ anoxically generated magnetite on steel. *J. Hazard. Mater. 147* (2007), 726–731.

[45] S. El Aamrani, J. Gimenez, M.Rovira, F. Seco, M. Grive, J. Bruno, L. Duro, J. de Pablo, A spectroscopic study of uranium(VI) interaction with magnetite, *Appl.Surf. Sci., 253* (2007) 8794–8797.

[46] T. Missana, C.Maffiotte,M. García-Gutiérrez , Surface reactions kinetics between nanocrystalline magnetite and uranyl , *Journal of Colloid and Interface Science, 261* (2003) 154–160. doi:10.1016/S0021-9797(02)00227-8.

[47] R. Leal, M. Yamaura, Equilibrium adsorption isotherm oF U(VI) aT pH4 and pH5 onto synthetic magnetite nanoparticles, In: Proc. of*2009 International Nuclear Atlantic Conference - INAC 2009, Rio de Janeiro,RJ, Brazil, September 27- October 2, 2009, Brazilian association of nuclear energy– ABEN, ISBN:* 978-85-99141-03-8.

[48] Manning, B.A.,Fendorf, S.E.,Bostick, B.,Suarez, D.L. (2002) Arsenic(III) oxidation and arsenic(V) adsorption reactions on synthetic birnessite, *Environmental Science and Technology*, 36 (2002) 976-981. doi: 10.1021/es0110170.

[49] J.B.Samuel, J.A.Stanley, D.P.Roopha, G. Vengatesh, J.Anbalagan, S.K. Banu, M.M. Aruldhas, Lactational hexavalent chromium exposure-induced oxidative stress in rat uterus is associated with delayed puberty and impaired gonadotropin levels , *Human and Experimental Toxicology, 30* (2011), 91-101. doi: 10.1177/0960327110364638.

[50] H. Catalette, J. Dumonceau, Ph. Ollar, Sorption of cesium, barium and europium on magnetite, *Journal of Contaminant Hydrology, 35* (1998) 151–159.

[51] W. Stumm, Chemistry of the Solid–Water Interface, Wiley Interscience, 1992, New York.

[52] M. Todorovic , S.K.Milonic, J.J. Comor, I.J. , I.J. Gal, Kinetics of Cs-137 sorption on natural magnetite, Radiation protection-selected topics, In : Proc. Int. Radiation Protection Symp., Dubrovnik, Yugoslavia, 1989, pp. 654-659, Boris Kidric (now VINCA) Institute of Nuclear Sciences, Belgrade.

[53] N. Marmier , F. Fromage, Sorption of Cs (I) on Magnetite in the Presence of Silicates, *Journal of Colloid and Interface Science, 223* (2000) 83–88.

[54] N. Marmier, A. Delisée, F. Fromage, Surface complexation modeling of Yb(III) and Cs(I) sorption on silica, *Journal of Colloid and Interface Science 212* (1999) 228-233.

[55] V. Philippini , A.Naveau, H. Catalette , S. Leclercq, Sorption of silicon on magnetite and other corrosion products of iron, *Journal of Nuclear Materials 348* (2006) 60–69. doi:10.1016/j.jnucmat.2005.09.002.

[56] P. Hu, X.Yin , L.Zhao , D.Li, Sorption of $Eu^{3+}$ onto nano-size silica-water interfaces, *Science in China Ser. D Earth Sciences,* 48 (2005) 1942-1948.

[57] M. M. Amin, A.Khodabakhshi, M. Mozafari, B. Bina, S. Kheiri, Removal of Cr(VI) from simulated electroplating wastewater by magnetite nanoparticles, *Env. Eng. Mngt J.,9* (2010) 921-927.

[58] K. Selvi, S. Pattabhi, K. Kadirvelu, Removal of Cr (VI) from aqueous solution by adsorption onto activated carbon, *Bioresource Technology,* 80 (2001) 87-89.

[59] P.Yuan, M.Fan , D.Yang, H. He, D.Liu, A.Yuan, J.Zhu, T.Chen, (2009), Montmorillonite-supported magnetite nanoparticles for the removal of hexavalent chromium $Cr(VI)$ from aqueous solutions, *J.Haz. Mat., 166* (2010) 821-829.

[60] N. Zhang, L.S. Lin, D.C. Gang, Adsorptive selenite removal from water using iron-coated GAC adsorbents, *Water Res. 42* (2008) 3809–3816.

[61] N. Belzile, Y.W.Chen, C.Y.Lang, M. Wang, The competitive role of organic carbon and dissolved sulfide in controlling the distribution of mercury in freshwater lake sediments, *Sci. Tot. Environ. 405* (2008) 226–238.

[62] Y.W.Chen, M.D. Zhou, J. Tong, N.Belzile, Application of photochemical reactions of Se in natural waters by hydride generation atomic fluorescence spectrometry, *Anal. Chim. Acta, 545* (2005) 142–148.

[63] M. Martınez , J. Gimenez , J. de Pablo, M. Rovira, L. Duro , Sorption of selenium(IV) and selenium(VI) onto magnetite , Applied Surface Science 252 (2006) 3767–3773. doi:10.1016/j.apsusc.2005.05.067 .

[64] A. El Hatimi, M. Martınez, L. Duro, J. de Pablo, Sorption of selenium onto natural iron oxides, Goldschmidt 2000, *J. Conf. Abs. 5* (2000) 380.

[65] N. Jordan, C. Lomenech, N. Marmier, E. Giffaut, J-J. Ehrhardt, Sorption of selenium (IV) onto magnetite in the presence of silicic acid, *Journal of Colloid and Interface Science 329* (2009) 17–23. doi:10.1016/j.jcis.2008.09.052.

[66] N. Jordan, N. Marmier, C. Lomenech, E. Giffaut, J-J. Ehrhardt, Sorption of silicates on goethite, hematite, and magnetite: Experiments and modelling , J*ournal of Colloid and Interface Science, 312,* (2007) 224-229. doi:10.1016/j.jcis.2007.03.053.

[67] T.S.Y. Choong , T.G. Chuaha, Y. Robiaha, F.L. Gregory Koaya, I. Azni , Arsenic toxicity, health hazards and removal techniques: An Overview, *Desalination 217* (2007) 139–166. doi:10.1016/j.desal.2007.01.015.

[68] G.A. Cutter, Kinetic controls on metalloid speciation in sea water, *Marine Chem., 40* (1992) 65–80.

[69] M. Bissen, F.H. Frimmel, Arsenic-A review. Part I: Occurrence, toxicity, speciation, mobility, *Acta Hydroch. Hydrob. 31* (2003) 9-18. doi: 10.1002/aheh.200390025.

[70] M.L. Pierce, C.B. Moore, Adsorption of arsenite and arsenate on amorphous iron hydroxide. *Water Res. 15* (1982) 1247-1253(1982).

[71] K.P. Raven, A. Jain, R.H. Loeppert, Arsenite and arsenate adsorption on ferrihydrite: Kinetics, equilibrium, and adsorption envelopes. *Environ. Sci. Technol.,. 32* (1998) 344 - 349.

[72] S. Yean, L. Cong, C.T. Yavuz, J.T. Mayo, W.W. Yu, A.T. Kan, V.L. Colvin, M.B. Tomson, Effect of magnetite particle size on adsorption and Desorption, of arsenite and arsenate, *J. Mater. Res.,* 20 (2005) 3255-3264. doi: 10.1557/JMR.2005.0403.

[73] K. Fukushi, T. Sato: Using a surface complexation model to predict the nature and stability of nanoparticles. *Environ. Sci. Technol.* **39** (2005) 1250-1256. doi: 10.1021/es0491984.

[74] T. Nagata, K. Fukushi, Prediction of iodate adsorption and surface speciation on oxides by surface complexation modeling, *Geochimica et Cosmochimica Acta*, 74 (2010) 6000-6013. doi:10.1016/j.gca.2010.08.002.

[75] J.T. Mayo, C. Yavuz, S. Yean, L. Cong, H. Shipley, W. Yu, J. Falkner, A. Kan, M. Tomson, V.L. Colvin, The effect of nanocrystalline magnetite size on arsenic removal, *Science and Technology of Advanced Materials,* 8 (2007) 71–75. doi:10.1016/j.stam.2006.10.005.

[76] D. Mohan, C. U. Pittman Jr., Arsenic removal from water/wastewater using adsorbents-A critical review, *Journal of Hazardous Materials,* 142 (2007) 1–53. doi:10.1016/j.jhazmat.2007.01.006.

[77] S. K. R. Yadanaparthi, D.Graybill, R. von Wandruszka, Adsorbents for the removal of arsenic, cadmium, and lead from contaminated waters, *Journal of Hazardous Materials,* 171 (2009) 1–15. doi:10.1016/j.jhazmat.2009.05.103.

[78] E.O. Kartinen, C.J.Martin, An overview of arsenic removal processes, *Dessalinaion, 103* (1995) 79-88.

[79] G.Dobby, J. A.Finch, Capture of mineral particles in a high gradient magnetic field, *Powder Technology. 17* (1977) 73- 82.

[80] J. A. Oberteuffer, Engineering development of High Gradient Magnetic Separators, *IEEE Trans. Magn. MAG-12.5* (1976), 444–449.

[81] Gerber, R., R. R. Birss, *High Gradient Magnetic Separation*, Research Studies Press, London, 1983.

[82] J. Svoboda, Application of High-Gradient Magnetic Separation, *Magnetic Separation News, 2* (1986) 51-68.

[83] A. D. Ebner, J. A. Ritter, L. Nufiez, High-Gradient Magnetic Separation for the Treatment of high-level radioactive wastes, *Separation Science and Technology, 34* (1999) 1333-1350.

[84] T. Oka, H. Kanayama, S. Fukui, J. Ogawa, T. Sato, M. Ooizumi, T. Terasawa, Y. Itoh, R. Yabuno, Application of HTS bulk magnet system to the magnetic separation techniques for water purification, *Physica C,* 468 (2008) 2128–2132.

[85] G.D., Moeser, K.A. Roach, W. H. Green, T. Alan Hatton, Hhigh-gradient magnetic separation of coated magnetic nanoparticles, *AIChE J.,* 50 (2004) 2835-2848.

[86] C. T.Yavuz, A. Prakash, J.T.Mayo, V.L.Colvin, Magnetic separations: From steel plants to biotechnology, *Chemical Engineering Science,* 64 (2009) 2510 – 2521. doi:10.1016/j.ces.2008.11.018.

[87] G. Mariani, M. Fabbri, F. Negrini, P. L.Ribani, High-Gradient Magnetic Separation of pollutant from wastewaters using permanent magnets, *Separation and Purification Technology* 72 (2010) 147–155. doi:10.1016/j.seppur.2010.01.017.

[88] V.K. Gupta, V.K. Saini, N. Jain, Adsorption of As(III) from aqueous solutions by iron oxide-coated sand, *Journal of Colloid and Interface Science,* 288 (2005) 55–60. doi:10.1016/j.jcis.2005.02.054.

[89] S. Kundu, A.K. Gupta, Analysis and modeling of fixed bed column operations on As(V) removal by adsorption onto iron oxide-coated cement (IOCC), *Journal of Colloid and Interface Science,* 290 (2005) 52–60. doi:10.1016/j.jcis.2005.04.006.

[90] I.Katsoyiannis, A. Zouboulis, H.Althoff, H.Bartel, As(III) removal from groundwaters using fixed-bed upflow bioreactors, *Chemosphere* 47 (2002) 325–332. PII: S00 4 5-6 5 35 (0 1 )0 03 0 6-X.

[91] T. S.Singh, K.K. Pant, Experimental and modelling studies on fixed bed adsorption of As(III) ions from aqueous solution, Separation and Purification Technology 48 (2006) 288–296. doi:10.1016/j.seppur.2005.07.035

[92] R. L. Kochen, J. D. Navratil, Method for Regenerating Magnetic Polyamine-Epichlorohydrine Resin, United States Patent 5,652,190 (July 29, 1997)

[93] W.Zhang, P.Singh, E.Paling, S.Delides, Arsenic removal from contaminated water by natural iron ores, *Minerals Engineering*, 17 (2004) 517–524, doi:10.1016/j.mineng.2003.11.020.

[94] J. Hu, I. M.C. Lo, G. Chen, Comparative study of various magnetic nanoparticles for Cr(VI) removal, *Separation and Purification Technology*, 56 (2007) 249–256. doi:10.1016/j.seppur.2007.02.009.

[95] J. Hristov, Magnetic field assisted fluidization-A unified approach.Part 1. Fundamentals and relevant hydrodynamics, *Reviews in Chemical Engineering, 18* (2002)295-509

[96] J.Hristov, Magnetic field assisted fluidization-A unified approach. Part 5. A Hydrodynamic Treatise on Liquid-solid Fluidized Beds, *Reviews in Chemical Engineering, 22* (2006) 195-377.

[97] J.Hristov, Magnetic field assisted fluidization-A unified approach. Part 7. Mass Transfer: Chemical reactors, basic studies and practical implementations thereof, *Reviews in Chemical Engineering, 25* (2009) 1-254.

[98] J.Hristov, L. Fachikov, Overview of Separations Performed by Magnetically Assisted Fluidized Beds with a special emphasis to waste stream cleaning, *China Particuology,5* (2007) 11-18. doi:10.1016/j.cpart.2007.01.010.

## Authors' information

[1]Department of Inorganic and Electrochemical Technologies,
University of Chemical technology and Metallurgy, Sofia, Bulgaria
[2]Second author affiliation.
Department of Inorganic and Electrochemical Technologies,
University of Chemical technology and Metallurgy, Sofia, Bulgaria
[3]Third author affiliation.
Department of Chemical engineering,
University of Chemical technology and Metallurgy, Sofia, Bulgaria

1. **Ms. Tanya Petrova** is a PhD student Department of Inorganic and Electrochemical Technologies. Bs and Ms in material science (leather technology). The research interests are on species deposition on magnetic solids and applications of magnetic fields.

2. **Ludmil Fachikov,** PhD ( Corrison Science) is associate professor in Department of Inorganic and Electrochemical Technologies. man research interests are in metal finishing ad corrosion protection.
e-mail : fachikov@uctm.edu

3. **Jordan Hristov** is associate professor in Department of Chemical Engineering University of Chemical Technology and Metallurgy (UCTM) 1756 Sofia, 8 Kl. Ohridsky Blvd., Bulgaria,
e-mail: jordan.hristov@mail.bg; hristovmeister@gmail.com
website : http://hristov.com/jordan